# THE ROLE OF LEGAL FRAMEWORKS IN SHAPING ETHICAL ARTIFICIAL INTELLIGENCE USE IN CORPORATE GOVERNANCE

Shahmar Mirishli∗

**Abstract**

*This article examines the evolving role of legal frameworks in shaping ethical artificial intelligence (AI) use in corporate governance. As AI systems become increasingly prevalent in business operations and decision-making, there is a growing need for robust governance structures to ensure their responsible development and deployment. Through analysis of recent legislative initiatives, industry standards, and scholarly perspectives, this paper explores key legal and regulatory approaches aimed at promoting transparency, accountability, and fairness in corporate AI applications. It evaluates the strengths and limitations of current frameworks, identifies emerging best practices, and offers recommendations for developing more comprehensive and effective AI governance regimes. The findings highlight the importance of adaptable, principle-based regulations coupled with sector-specific guidance to address the unique challenges posed by AI technologies in the corporate sphere.*

**Keywords:** *artificial intelligence, corporate governance, AI ethics, AI regulation, algorithmic accountability, legal frameworks, data protection, corporate law, technology policy, regulatory compliance, risk management, digital transformation.*

## I. Introduction

The rapid advancement and integration of artificial intelligence (AI) into corporate operations has fundamentally altered the landscape of business decision-making and governance. As AI systems increasingly influence critical aspects of corporate function, from strategic planning to operational execution, the need for robust legal frameworks to guide their ethical development and deployment has become paramount. This article examines the complex interplay between emerging AI technologies, corporate governance structures, and legal regulatory frameworks.

The proliferation of AI in corporate settings presents both unprecedented opportunities and significant challenges. On one hand, AI technologies offer the potential for enhanced efficiency, data-driven insights, and innovative business models. On the other, they raise profound questions about accountability, transparency, and the potential for unintended consequences or biases that could significantly impact individuals and society at large.

The central question this paper addresses is: How can legal and regulatory frameworks effectively promote ethical AI use in corporate governance while fostering innovation? To explore this issue, we will examine several interconnected themes:

1. The evolving landscape of AI-specific regulations and their impact on corporate practices, including an analysis of the contrasting approaches taken by major jurisdictions such as the European Union and the United States.

2. The role of existing legal frameworks, particularly in areas such as data protection, antitrust, and corporate law, in governing AI applications and their adequacy in addressing novel challenges posed by AI technologies.

3. Industry-led initiatives and self-regulatory efforts to establish AI governance standards, evaluating their effectiveness and limitations in promoting responsible AI development and use.

---
∗ Lawyer

4. The challenges in balancing innovation with responsible AI development, exploring the tension between fostering technological progress and mitigating potential risks.

5. Emerging best practices and recommendations for comprehensive AI governance, drawing insights from current initiatives and scholarly perspectives.

By analyzing these topics, this paper aims to provide insights into the complex relationship between legal frameworks, corporate governance structures, and ethical AI development. The findings will contribute to ongoing discussions on how to design adaptive and effective governance regimes for rapidly evolving AI technologies in the business world.

As AI continues to transform corporate landscapes, the legal and regulatory frameworks governing its use must evolve in tandem. This article seeks to illuminate the current state of AI governance, identify key challenges and opportunities, and offer pathways for developing more robust and effective legal approaches to ensuring ethical AI use in corporate settings. In doing so, it aims to contribute to the crucial dialogue on how best to harness the transformative potential of AI while safeguarding fundamental rights, ethical principles, and the broader public interest.

**II. Legal Frameworks and Corporate AI Governance: Current Landscape and Emerging Approaches**

The rapid advancement and integration of artificial intelligence (AI) into corporate operations has necessitated a fundamental reevaluation of existing legal and regulatory frameworks. This section examines the evolving landscape of AI governance, focusing on key jurisdictions' approaches, the application of existing legal structures, and the emergence of industry-led initiatives. The complex interplay between these various elements presents both challenges and opportunities for the development of comprehensive and effective AI governance regimes.

*1. The European Union's Comprehensive Strategy*

The European Union has positioned itself at the forefront of AI regulation with its proposed Artificial Intelligence Act [1]. This legislation represents a significant departure from previous regulatory approaches, adopting a risk-based framework that categorizes AI systems based on their potential impact on individuals and society. The Act delineates four risk categories: unacceptable risk, high risk, limited risk, and minimal risk.

Systems deemed to pose unacceptable risks, such as those involving subliminal manipulation or social scoring by public authorities, are prohibited outright. This categorical prohibition reflects a strong precautionary approach, prioritizing the protection of fundamental rights over potential technological benefits. However, it also raises questions about the potential stifling of innovation in these areas and the challenges of defining and identifying "unacceptable" risks in a rapidly evolving technological landscape.

High-risk AI systems, which include applications in critical infrastructure, education, employment, and law enforcement, are subject to stringent requirements. These requirements encompass the implementation of risk management systems, high-quality data governance, detailed documentation and record-keeping, transparency and provision of information to users, human oversight, and robustness, accuracy, and cybersecurity measures. The comprehensive nature of these requirements reflects an attempt to address the multifaceted challenges posed by AI systems, but also raises concerns about the potential regulatory burden on businesses, particularly small and medium-sized enterprises.

The Act's extraterritorial scope means that it will impact any entity providing AI systems or services within the EU market, regardless of their geographical location. This broad reach could potentially lead to a "Brussels effect," whereby EU standards become de facto global norms for AI governance. The extraterritorial application of the Act raises complex questions of international law and regulatory jurisdiction, potentially leading to conflicts with other regulatory regimes and challenges in enforcement.

While the EU's approach provides a comprehensive framework for AI regulation, it is not without criticism. Some argue that the stringent requirements could stifle innovation, particularly for smaller companies that may struggle with compliance costs. Additionally, the categorization of risk levels may prove challenging in practice, given the rapidly evolving nature of AI technologies. The Act's focus on ex-ante regulation also raises questions about its ability to adapt to unforeseen technological developments and its potential impact on the EU's competitiveness in the global AI market.

### 2. The United States' Sectoral Approach

In contrast to the EU's comprehensive strategy, the United States has thus far favored a more fragmented, sector-specific approach to AI regulation. This approach reflects the US's traditionally more market-driven regulatory philosophy and its emphasis on fostering innovation. The absence of overarching federal AI legislation has resulted in a patchwork of regulations and guidelines, with existing laws in areas such as consumer protection, antitrust, and civil rights being applied to AI-related issues.

The Federal Trade Commission (FTC) has emerged as a key player in AI governance in the US, asserting its authority to address unfair or deceptive practices involving AI under Section 5 of the FTC Act [5]. This approach leverages existing consumer protection frameworks to address AI-related issues, but it may be limited in its ability to comprehensively address the unique challenges posed by AI technologies.

The recent White House Executive Order on AI, issued in October 2023, marks a significant step towards a more coordinated national strategy [19]. The Order addresses various aspects of AI governance, including safety and security, privacy protection, equity and civil rights, consumer and worker protection, and the promotion of innovation and competition. It directs federal agencies to develop AI risk management frameworks, establishes new standards for AI safety and security, and aims to strengthen privacy guidance for federal agencies.

Key provisions of the Executive Order include:

1. Requiring developers of powerful AI systems to share safety test results with the U.S. government. This provision raises important questions about the balance between transparency and the protection of proprietary information, as well as potential challenges in defining and measuring "safety" in the context of AI systems.

2. Directing the National Institute of Standards and Technology (NIST) to develop guidelines for extensive red team testing of AI systems. While this approach aims to enhance the robustness and security of AI systems, it also raises questions about the standardization of testing methodologies and the potential for such tests to be gamed or circumvented.

3. Evaluating how agencies collect and use commercially available information containing personal data. This provision reflects growing concerns about data privacy in the age of AI, but its effectiveness will depend on the development of clear guidelines and enforcement mechanisms.

4. Providing guidance to landlords, Federal benefits programs, and federal contractors to prevent AI algorithms from exacerbating discrimination. This approach recognizes the potential for AI systems to perpetuate or exacerbate existing societal biases, but its success will depend on the development of effective methods for detecting and mitigating algorithmic bias.

5. Developing principles and best practices to mitigate the negative impacts and maximize the benefits of AI for workers. This provision acknowledges the potential disruptive effects of AI on the labor market, but the development of effective policies in this area will require careful balancing of worker protection with the need for economic innovation and competitiveness.

6. Catalyzing AI research through a pilot of the National AI Research Resource. While this initiative aims to democratize access to AI research resources, it also raises questions about the appropriate role of government in driving technological innovation and the potential for such resources to be misused or exploited.

The US approach offers greater flexibility and potentially allows for faster adaptation to technological changes. However, it also risks creating regulatory gaps and inconsistencies across sectors and states. This fragmented landscape poses challenges for corporations operating nationwide, as they must navigate varying requirements and standards. The lack of a comprehensive federal AI law may also put the US at a disadvantage in shaping global AI governance norms, potentially ceding this role to other jurisdictions with more cohesive regulatory approaches.

*3. Comparative Analysis of EU and US Approaches*

The contrasting approaches of the EU and US to AI regulation reflect fundamental differences in regulatory philosophy and risk assessment. Smuha [17] highlights several key distinctions:

1. Scope: The EU's approach is comprehensive and applies across sectors, while the US approach is more fragmented and sector-specific. This difference reflects broader divergences in regulatory philosophy between the two jurisdictions, with the EU favoring a more centralized, harmonized approach and the US preferring a more decentralized, market-driven model.

2. Risk Assessment: The EU explicitly categorizes AI systems based on risk levels, while the US approach is more implicit in its risk assessment. The EU's approach provides greater clarity and predictability for businesses but may be less flexible in dealing with novel AI applications that don't fit neatly into predefined risk categories.

3. Prescriptiveness: The EU AI Act provides detailed requirements for high-risk AI systems, whereas the US approach generally relies on broader principles and existing legal frameworks. This difference raises questions about the relative merits of rules-based versus principles-based regulation in the context of rapidly evolving technologies.

4. Innovation Focus: The US approach places a stronger emphasis on fostering innovation, while the EU approach prioritizes risk mitigation. This reflects different societal values and priorities, with the US traditionally favoring a more laissez-faire approach to technological development and the EU placing greater emphasis on precautionary principles.

5. Enforcement: The EU proposes a dedicated enforcement mechanism for AI regulation, while the US relies more on existing regulatory bodies and their enforcement powers. The effectiveness of these different enforcement approaches in ensuring compliance and protecting individual rights remains to be seen.

These differences create challenges for multinational corporations, which must navigate varying regulatory requirements across jurisdictions. The potential for regulatory arbitrage and forum shopping by AI developers and users is a significant concern, as is the risk of conflicting compliance obligations for companies operating in both markets.

However, these divergent approaches also provide opportunities for regulatory learning and potential future harmonization of approaches. The coexistence of different regulatory models allows for comparative analysis of their effectiveness, potentially informing the development of more refined and universally applicable AI governance frameworks in the future.

The global nature of AI development and deployment underscores the need for international cooperation and coordination in AI governance. Efforts towards developing common principles and standards, such as those undertaken by the OECD and other international organizations, will be crucial in bridging the gaps between different regulatory approaches and fostering a more coherent global AI governance regime.

*4. Industry-Led Initiatives and Self-Regulation*

In response to the evolving regulatory landscape and the need for flexible governance mechanisms, many corporations and industry groups have launched self-regulatory initiatives. These efforts aim to establish voluntary standards and best practices for ethical AI development and use, reflecting a recognition of the potential reputational and operational risks associated with unethical or irresponsible AI practices.

Notable examples include the Partnership on AI, which brings together leading tech companies, academic institutions, and civil society organizations to collaborate on responsible AI practices. This multi-stakeholder approach reflects an understanding of the complex and interdisciplinary nature of AI governance challenges, but also raises questions about potential conflicts of interest and the ability of industry-led initiatives to adequately protect public interests.

The OECD Principles on Artificial Intelligence, adopted in 2019, represent a significant multilateral effort to establish guidelines for responsible AI development [14]. These principles emphasize:

1. Inclusive growth, sustainable development, and well-being
2. Human-centered values and fairness
3. Transparency and explainability
4. Robustness, security, and safety
5. Accountability

These principles have been influential in shaping both corporate practices and national AI strategies. However, their voluntary nature raises questions about enforceability and consistency. The lack of binding mechanisms to ensure compliance with these principles may limit their effectiveness in driving meaningful change in corporate AI practices.

Floridi et al. [10] propose seven essential factors for designing AI for social good:
1. Falsifiability and incremental deployment

2. Safeguards against the manipulation of predictors
3. Receiver-contextualized intervention
4. Receiver-contextualized explanation and transparent purposes
5. Privacy protection and data subject consent
6. Situational fairness
7. Human-friendly semanticisation

These factors provide a framework for corporations to consider when developing and deploying AI systems, particularly in contexts where the social impact of AI is significant. However, the practical implementation of these factors in complex AI systems remains challenging, and there is a need for more concrete guidance on how to operationalize these principles in diverse organizational contexts.

Other significant industry-led initiatives include the IEEE Global Initiative on Ethics of Autonomous and Intelligent Systems, the AI Now Institute, OpenAI, and AI4People. While these initiatives demonstrate a commitment to responsible AI development, their voluntary nature raises questions about enforceability and consistency. Critics argue that self-regulation alone is insufficient to address the full scope of AI governance challenges, highlighting the need for complementary regulatory frameworks.

The effectiveness of industry-led initiatives in promoting ethical AI use varies. While some corporations have made significant strides in implementing robust AI governance frameworks, others have been criticized for paying lip service to ethical principles without substantive changes to their practices [16]. This variability in implementation raises questions about the reliability of self-regulatory approaches and the need for external oversight and enforcement mechanisms.

The strengths of industry-led initiatives include flexibility to adapt quickly to technological changes, leveraging of industry expertise, and the potential to foster innovation while addressing ethical concerns. However, they also face weaknesses such as lack of enforcement mechanisms, potential conflicts of interest, and the risk of fragmentation leading to inconsistent standards across the industry.

The relationship between industry-led initiatives and formal regulatory frameworks is an area of ongoing debate. While some argue that self-regulation can complement and inform formal regulation, others contend that it may be used as a means to preempt or weaken more stringent regulatory measures. The appropriate balance between self-regulation and formal regulation in AI governance remains a key challenge for policymakers and industry leaders alike.

*5. Application of Existing Legal Frameworks*

While AI-specific regulations are still evolving, existing legal frameworks play a crucial role in governing corporate AI practices. Data protection regulations, particularly the EU's General Data Protection Regulation (GDPR), have emerged as key tools for governing AI systems [8]. The GDPR's principles of data minimization, purpose limitation, and the right to explanation for automated decisions have significant implications for AI development and deployment.

The GDPR includes several key provisions relevant to AI governance:

1. Data Minimization (Article 5(1)(c)): Personal data must be adequate, relevant, and limited to what is necessary in relation to the purposes for which they are processed. This principle poses significant challenges for AI systems that often require large datasets for training and operation. The tension between data minimization and the data-intensive nature of many AI applications raises questions about the compatibility of current data protection frameworks with the realities of AI development.

2. Purpose Limitation (Article 5(1)(b)): Personal data must be collected for specified, explicit, and legitimate purposes and not further processed in a manner that is incompatible with those purposes. This principle can be problematic for AI systems that may find new, unforeseen patterns or uses for data. The inherent unpredictability of some AI systems, particularly those using machine learning techniques, challenges traditional notions of purpose specification in data protection law.

3. Right to Explanation (Articles 13-15, 22): Individuals have the right to meaningful information about the logic involved in automated decision-making. This right poses challenges for complex AI systems, particularly those using deep learning techniques, where the decision-making process may not be easily interpretable. The practical implementation of this right remains a subject of debate, with questions about what constitutes a "meaningful" explanation in the context of complex AI systems.

4. Data Protection Impact Assessments (Article 35): Organizations must conduct impact assessments for high-risk data processing activities, which would include many AI applications. The application of DPIAs to AI systems raises questions about the appropriate methodologies for assessing AI-specific risks and the capacity of organizations to conduct meaningful assessments of complex AI systems.

5. Data Protection by Design and by Default (Article 25): Organizations must implement appropriate technical and organizational measures to implement data protection principles and safeguard individual rights. This principle aligns with the concept of "AI ethics by design," but its practical implementation in AI development processes remains challenging.

The application of GDPR principles to AI systems raises complex legal questions. For instance, the requirement to provide "meaningful information about the logic involved" in automated decisions poses particular challenges for deep learning systems whose decision-making processes may not be easily interpretable [5]. The tension between the GDPR's data minimization principle and the data-hungry nature of many AI systems presents another challenge for corporate compliance [7].

In the United States, sector-specific laws have implications for AI systems:

1. Fair Credit Reporting Act (FCRA): Regulates the collection, dissemination, and use of consumer information, including in automated decision-making systems used in credit, employment, and insurance contexts. The application of FCRA to AI-driven credit scoring systems raises questions about the adequacy of traditional notions of "credit worthiness" in the age of big data and machine learning.

2. Equal Credit Opportunity Act (ECOA): Prohibits discrimination in credit transactions, which has implications for AI systems used in lending decisions. The use of AI in credit decisions raises complex questions about algorithmic bias and the potential for AI systems to perpetuate or exacerbate existing patterns of discrimination in lending practices.

3. Americans with Disabilities Act (ADA): Requires equal access to public accommodations, which could apply to AI-powered services and platforms. The application of the ADA to AI systems raises novel questions about what constitutes "reasonable accommodation" in the context of AI-driven services and the potential for AI to both enhance and hinder accessibility for individuals with disabilities.

4. Title VII of the Civil Rights Act: Prohibits employment discrimination, which is relevant to AI systems used in hiring and promotion decisions. The use of AI in employment decisions raises concerns about algorithmic bias and the potential for AI systems to perpetuate or exacerbate existing patterns of workplace discrimination.

These laws prohibit discrimination and often require explanations for adverse decisions, principles that are directly relevant to AI-driven decision-making processes. However, their application to complex AI systems raises questions about the adequacy of traditional anti-discrimination frameworks in addressing the unique challenges posed by algorithmic decision-making.

Antitrust law is another area where existing legal frameworks are being applied to AI-related issues. The concentration of AI capabilities among a few large tech companies has raised concerns about market dominance and potential anti-competitive practices. Key issues include:

1. Data Monopolies: The vast amounts of data held by large tech companies can create barriers to entry for new competitors in AI markets. This raises questions about the adequacy of traditional antitrust frameworks in addressing data-driven market power and the potential need for new approaches to data regulation and competition policy.

2. Network Effects: AI-powered platforms can benefit from strong network effects, potentially leading to winner-take-all markets. The self-reinforcing nature of AI-driven network effects challenges traditional antitrust analysis and may require new approaches to market definition and assessment of market power.

3. Algorithmic Collusion: There are concerns that AI systems could potentially engage in tacit collusion, even without explicit instructions to do so. This raises novel questions about the application of antitrust laws to autonomous systems and the potential need for new legal frameworks to address AI-driven market manipulation.

4. Merger Control: The acquisition of AI startups by large tech companies has raised questions about how to assess the competitive implications of such mergers. Traditional merger analysis frameworks may be inadequate in capturing the long-term competitive impacts of AI-related acquisitions, particularly given the rapid pace of technological change in the AI field.

Traditional antitrust frameworks may need to be adapted to address these unique characteristics of AI markets [15]. The intersection of AI and antitrust law represents a rapidly evolving area of legal scholarship and practice, with significant implications for corporate strategy and regulatory policy.

*6. Corporate Governance and AI*

The integration of AI into corporate decision-making processes raises novel questions in corporate law and governance. As AI systems take on more significant roles in areas traditionally reserved for human judgment, existing legal frameworks for corporate accountability and fiduciary duty may need to be reevaluated.

Armour and Eidenmueller [2] propose the concept of "self-driving corporations," where AI systems play an increasingly central role in corporate decision-making. This raises complex legal questions about accountability, fiduciary duty, and the role of human oversight in AI-driven corporate governance. Key issues include:

1. Board Oversight: Corporate boards may need to develop new competencies to effectively oversee AI systems and their risks. This may require changes in board composition and training to ensure adequate understanding of AI technologies and their implications. The traditional role of the board as the ultimate decision-making authority in a corporation may need to be reconsidered in light of AI systems that can process vast amounts of data and make complex decisions at speeds far beyond human capability.

2. Fiduciary Duty: The use of AI in corporate decision-making may complicate traditional understandings of directors' and officers' fiduciary duties. Questions arise about the extent to which reliance on AI systems affects the duty of care and the business judgment rule. For instance, if an AI system makes a decision that leads to corporate losses, to what extent can directors claim protection under the business judgment rule if they relied on the AI's recommendation? The concept of "informed decision-making" may need to be redefined to account for the complexities of AI-driven decision processes.

3. Disclosure Requirements: Companies may face new obligations to disclose material information about their use of AI systems to shareholders and regulators. This could include information about the role of AI in strategic decision-making, AI-related risks and risk management processes, and potential impacts of AI on financial performance and business models. The challenge lies in determining what constitutes "material" information in the context of AI use, given the often opaque nature of AI decision-making processes.

4. Liability: The use of AI in corporate decision-making raises complex questions about liability allocation in cases where AI-driven decisions lead to harm. Traditional concepts of corporate liability may need to be reconsidered in light of autonomous AI systems. For example, if an AI system makes a decision that causes harm to stakeholders, how should liability be allocated between the corporation, its directors, the AI developers, and potentially the AI system itself? The concept of "AI personhood" and its implications for corporate law is an area of growing scholarly debate.

5. Shareholder Rights: The use of AI in corporate governance may impact shareholder rights and engagement. AI could potentially be used to enhance shareholder participation in corporate decision-making, for instance through AI-powered voting systems or AI-assisted shareholder proposals. However, this also raises concerns about the potential for AI to be used to manipulate shareholder sentiment or to create information asymmetries between different classes of shareholders.

6. Ethical Considerations: Corporations need to consider how to embed ethical considerations into their AI governance frameworks. This includes developing AI ethics committees or advisory boards, implementing ethical guidelines for AI development and deployment, ensuring diversity and inclusivity in AI teams and datasets, and considering the broader societal impacts of AI systems. The challenge lies in translating broad ethical principles into concrete operational guidelines and in ensuring that ethical considerations are not sidelined in pursuit of efficiency or profit.

These issues highlight the need for a fundamental reevaluation of corporate governance structures and practices in the age of AI. Companies will need to develop robust governance frameworks that address the unique challenges posed by AI while ensuring compliance with existing legal obligations. This may require new legal and regulatory approaches that can accommodate the dynamic and often unpredictable nature of AI systems.

*7. Challenges in AI Governance*
Despite the proliferation of AI governance initiatives, significant challenges remain in creating effective legal and regulatory frameworks. These challenges include:

1. Balancing Innovation and Regulation: Striking the right balance between fostering innovation and ensuring responsible development is a key challenge. Overly prescriptive regulations risk stifling technological progress, while inadequate oversight could lead to harmful outcomes [13]. The rapid pace of AI development makes it difficult for regulators to keep up, potentially leading to a regulatory lag that could have significant societal implications.

2. Explainability and Interpretability: The "black box" nature of many AI systems, particularly those based on deep learning, poses challenges for transparency and accountability. Developing clear standards for AI explainability and interpretability, particularly for complex machine learning models, remains a significant hurdle [5]. This challenge is particularly acute in high-stakes domains such as healthcare, criminal justice, and financial services, where the ability to understand and explain AI decisions is crucial for public trust and legal compliance.

3. Global Harmonization: The global nature of AI development and deployment necessitates greater international cooperation in governance. However, achieving harmonization across different legal and regulatory systems poses significant challenges [17]. Differences in cultural values, legal traditions, and economic priorities make it difficult to establish universally accepted AI governance norms. The potential for "regulatory arbitrage," where AI developers choose to operate in jurisdictions with less stringent regulations, is a significant concern.

4. Rapid Technological Advancement: The pace of AI development often outstrips regulatory efforts, creating potential gaps and inconsistencies in governance frameworks. Regulatory approaches need to be flexible enough to adapt to rapidly evolving technologies [18]. This may require new approaches to regulation, such as "regulatory sandboxes" or "adaptive regulation" models that can evolve in tandem with technological developments.

5. Long-term Societal Impacts: Assessing and mitigating the long-term societal impacts of widespread AI adoption remains a complex challenge. Current legal frameworks may not adequately address these potential impacts [6]. Issues such as AI's effect on employment, social cohesion, and democratic processes require careful consideration and may necessitate novel policy approaches.

6. Data Governance: The data-intensive nature of AI systems creates tensions with data protection principles and raises questions about data ownership, access, and control [8]. The global flow of data necessary for AI development and deployment challenges traditional notions of data sovereignty and jurisdictional authority.

7. Algorithmic Bias and Fairness: Ensuring fairness and preventing discrimination in AI systems remains a significant challenge, particularly given the potential for AI to perpetuate or exacerbate existing societal biases [10]. Developing robust methods for detecting and mitigating bias in AI systems, while also ensuring that these methods themselves do not introduce new forms of bias, is an ongoing area of research and policy development.

8. AI Safety and Security: Ensuring the safety and security of AI systems, particularly as they become more autonomous and influential in critical domains, is a growing concern [19]. This includes addressing issues such as AI system vulnerability to adversarial attacks, the potential for AI systems to be used maliciously, and the challenges of ensuring AI alignment with human values and goals.

9. AI and Employment: The potential impact of AI on employment and the nature of work raises complex policy challenges, including the need for workforce reskilling and potential changes to labor laws [12]. The potential for AI to automate a wide range of tasks raises questions about the future of work and the need for new social and economic models to address potential job displacement.

10. AI and Intellectual Property: The use of AI in creative processes and innovation raises new questions about intellectual property rights and the nature of authorship and inventorship [5]. Traditional intellectual property frameworks may be inadequate to address issues such as AI-generated inventions or creative works, potentially requiring fundamental reconsideration of intellectual property law.

Addressing these challenges will require ongoing collaboration between policymakers, industry leaders, researchers, and civil society to develop nuanced, context-specific approaches to AI ethics and risk management. The complexity and interconnectedness of these issues underscore the need for a holistic approach to AI governance that considers not only technical and legal aspects but also broader societal and ethical implications.

*8. Future Directions in AI Governance*

As the field of AI continues to evolve rapidly, several key trends and potential future directions in AI governance are emerging:

1. Adaptive Regulation: There is growing recognition of the need for more flexible and adaptive regulatory approaches that can keep pace with technological change. This may include the development of "regulatory sandboxes" where new AI applications can be tested under controlled conditions, or the use of "soft law" instruments that can be more easily updated than traditional legislation.

2. Algorithmic Auditing: The development of standardized methods for auditing AI systems for bias, fairness, and other ethical considerations is likely to be a key focus of future governance efforts. This may include the emergence of third-party auditing services and the development of technical standards for AI auditing.

3. AI Ethics by Design: There is increasing emphasis on incorporating ethical considerations into the AI development process from the outset, rather than treating ethics as an afterthought. This may lead to the development of new design methodologies and tools that embed ethical principles into AI systems at a fundamental level.

4. International Cooperation: Given the global nature of AI development and deployment, there is likely to be increased focus on international cooperation and the development of global governance frameworks for AI. This may include efforts to harmonize regulations across jurisdictions and the establishment of international bodies to oversee AI governance.

5. Sectoral Approaches: While there are efforts to develop overarching AI governance frameworks, there is also recognition of the need for sector-specific approaches that address the unique challenges and risks associated with AI use in particular domains such as healthcare, finance, and criminal justice.

6. AI and Human Rights: There is growing attention to the implications of AI for human rights, and future governance frameworks are likely to place increased emphasis on ensuring that AI development and deployment respects and promotes human rights principles.

7. Participatory Governance: There are calls for more inclusive and participatory approaches to AI governance, involving a wider range of stakeholders in the development of policies and regulations. This may include greater use of public consultations, citizen juries, and other mechanisms for incorporating diverse perspectives into AI governance.

8. AI Literacy: As AI becomes increasingly pervasive, there is likely to be greater focus on enhancing AI literacy among policymakers, business leaders, and the general public. This may include the development of educational programs and public awareness campaigns about AI and its implications.

9. Governance of Advanced AI: As AI systems become more sophisticated and potentially approach or exceed human-level intelligence in certain domains, new governance challenges are likely to emerge. This may include considerations of AI agency, rights, and responsibilities, as well as the development of governance frameworks for potential future artificial general intelligence (AGI) or artificial superintelligence (ASI).

10. Interdisciplinary Approaches: Given the complex and multifaceted nature of AI governance challenges, there is likely to be increased emphasis on interdisciplinary approaches that bring together insights from computer science, law, ethics, social sciences, and other relevant fields.

These future directions highlight the dynamic and evolving nature of AI governance. As AI technologies continue to advance and their societal impacts become more pronounced, the legal and regulatory landscape will need to evolve in tandem. This will require ongoing dialogue, research, and collaboration among diverse stakeholders to ensure that AI governance frameworks are effective, ethical, and adaptable to future technological developments.

### III. Conclusion

The integration of artificial intelligence into corporate operations and decision-making processes presents unprecedented challenges to existing legal frameworks and corporate governance structures. This analysis reveals a complex landscape where regulatory approaches, industry initiatives, and established legal doctrines intersect, often struggling to keep pace with the relentless advancement of AI technologies.

The divergent regulatory strategies adopted by major jurisdictions, exemplified by the EU's comprehensive risk-based approach and the US's sectoral model, underscore the global struggle to balance innovation with responsible development. These contrasting approaches highlight the inherent tension between prescriptive regulation and adaptive governance in the face of rapidly evolving technologies.

Industry-led initiatives and self-regulatory efforts, while demonstrating a commitment to ethical AI development, raise critical questions about enforceability and potential conflicts of interest. The variability in the implementation and effectiveness of these voluntary measures underscores the need for more standardized approaches and potentially greater regulatory oversight.

The application of existing legal frameworks to AI-related issues reveals both the adaptability and limitations of current jurisprudence. Principles of data protection, antitrust law, and corporate governance are being stretched to their conceptual limits when applied to AI systems. This strain on established legal doctrines necessitates a fundamental reevaluation of core legal concepts such as liability, personhood, and fiduciary duty in the context of AI-driven decision-making.

The concept of "self-driving corporations" challenges traditional notions of corporate governance and director responsibilities. As AI systems assume increasingly central roles in corporate decision-making, the legal framework must evolve to address novel questions of accountability, oversight, and the appropriate balance between human judgment and machine intelligence.

Significant challenges persist in the realm of AI governance, including ensuring explainability and interpretability of AI systems, achieving global harmonization of governance approaches, and addressing the long-term societal impacts of widespread AI adoption. The rapid pace of technological advancement creates a persistent tension between the need for regulatory certainty and the imperative to foster innovation.

Looking ahead, the future of AI governance is likely to be characterized by more adaptive regulatory approaches, increased focus on algorithmic auditing, and the emergence of sector-specific governance frameworks. The growing recognition of AI's implications for fundamental rights and the need for more participatory governance approaches point towards a more nuanced and context-sensitive legal framework for AI.

In conclusion, the governance of AI in corporate contexts demands a paradigm shift in legal thinking and regulatory approach. It requires a delicate balance between fostering innovation and ensuring responsible development, between leveraging the transformative potential of AI and safeguarding fundamental rights and societal values. The legal framework must evolve to provide clear guidelines and robust protections while remaining flexible enough to accommodate the dynamic nature of AI technologies.

The challenge for corporate law and governance in the coming decades will be to develop a coherent and effective legal regime that can navigate the complexities of AI-driven corporate decision-making. This will necessitate not only legal and regulatory innovation but also a fundamental reimagining of the relationship between corporations, technology, and society. The ultimate goal must be to harness the power of AI to enhance corporate efficiency and innovation while ensuring that its development and deployment align with ethical principles and contribute to the broader public good. Achieving this balance will be crucial in shaping the future landscape of corporate law and governance in the age of artificial intelligence.